\documentclass[runningheads]{svmult}

\usepackage{makeidx}   
\usepackage{graphicx}  
\usepackage{subeqnar}  
\usepackage{multicol}  
\usepackage{physprbb}  
\makeindex             

\newcommand{\etab}{\mbox{\boldmath $\eta $}}
\newcommand{\Pib}{\mbox{\boldmath $\Pi $}}
\newcommand{\br}{{\bf r}}
\newcommand{\bp}{{\bf p}}
\newcommand{\bR}{{\bf R}}
\newcommand{\bq}{{\bf q}}
\newcommand{\qr}{{\bf q\cdot r}}
\newcommand{\qR}{{\bf q\cdot R}}
\newcommand{\barro}{\bar{\bar{\rho}}({\bf q})}

\newcommand{\sig}{\sigma_{xy}}
\newcommand{\n}{{p \over 2ps+1}}
\newcommand{\eeq}{\end{equation}}

\newcommand{\beq}{\begin{equation}}
\newcommand{\un}{2s + {1\over p}}

\begin{document}
\title*{Theories of the Fractional Quantum Hall Effect}
%
%
%
%
%
\author{R.Shankar}
\authorrunning{R.Shankar}
%
%
\institute{Yale University, New Haven CT 06520, USA }

\maketitle              

\begin{abstract}
This is an  introduction to the microscopic theories of the FQHE.
After a brief description of experiments, trial wavefunctions and
the physics they contain are discussed. This is followed by a
description of the hamiltonian approach, wherein one goes from the
electrons to the composite fermions by a series of
transformations. The theory is then compared to other theoretical
approaches and to experiment.
\end{abstract}

\section{Introduction}
In these lectures I will provide an introduction to some of the
theories of the Fractional Quantum Hall Effect (FQHE), for  a
non-FQHE person who is familiar with graduate quantum mechanics
and a little bit of second quantization. I will discuss only the
simplest phenomena to illustrate the basic ideas and also focus on
the parts of the subject I am most familiar with, since many
excellent reviews exist \cite{review1}, \cite{review2},
\cite{olle}. A rule I will bear in mind is that things should be
made as simple as possible but no simpler.
\section{The problem}
\begin{figure}[t]
\begin{center}
\includegraphics[width=.4\textwidth]{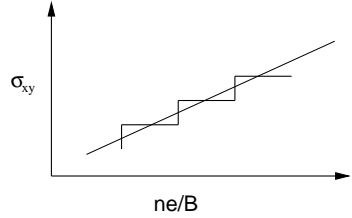}
\end{center}
\vspace*{-.7in} \caption[]{Schematic form of the Hall conductance:
the straight line is expected,  and the steps are seen. Professor
Glattli's lectures  discuss the data in
 great detail.} \label{hall}
\end{figure}

In the Hall effect, one takes a system of electrons in the $x-y$
plane, subject to a magnetic field in the $z$-direction. A current
density $j_x$ is driven in the $x$-direction and the voltage $V_y$
is measured in the $y$-directions. At equilibrium, an electric
field $E_y$ is set up in the $y$-direction which balances the
Lorentz force $v_xB$,
 and the Hall conductance is
 \beq
 \sig = {j_x\over E_y}={nev_x\over v_xB}={ne \over B}
 \eeq
 where $n$ is the number density.
 Thus we expect that $\sig$ will be linear in
 ${ne/B}$. Instead we find steps at some special values as in
 Figure   \ref{hall}.

\beq \sigma_{xy} = {e^2 \over 2\pi \hbar}\       {p \over
2ps+1}\equiv {e^2 \over 2\pi \hbar}       \nu \eeq

 \beq
 p=1,2 \ldots \ \ s=0,1,2\ldots \eeq

The case  $s=0$, $\nu =p$ corresponds to the Integer Quantum Hall
Effect (IQHE) discovered by   von Klitzing  {\em et al}
\cite{baron}. The case $s>0$ corresponding  to the FQHE was
discovered by Tsui {\em et al} \cite{fqhe-ex}. The case
 $p=1, \ \nu =1/(2s+1)$ was explained by
 Laughlin  \cite{laugh}
 with his famous trial state. The more general
  case $s\ge 0$, $p\ge 1, \nu ={p\over 2ps+1}$ was explained by Jain
  using the idea of composite fermions (CF)  \cite{jain-cf}.
(The case  $\nu = {p \over 2ps-1}$ is a  trivial extension.)
\subsection{What is special about these values of $\sig$?}
Note that where the straight line meets the  plateaus,

\begin{eqnarray}  \sigma_{xy}& =&     {e^2 \over 2\pi \hbar}
 {p \over 2ps+1}={ne\over B}
\\ {eBL^2\over 2\pi \hbar n L^2}&=& \un
\\ {\left [BL^2/({2\pi \hbar \over e})\right] \over nL^2}&=& {\Phi
/\Phi_0 \over N} = \un
\end{eqnarray}
where $\Phi_0={2\pi\hbar\over e}$ is a quantum of flux.  {\em At
these points, there are $\nu^{-1}=\un$} flux quanta per particle.
Thus IQHE has ${1\over p}$ flux quanta per electron. We will see
that this makes life easy, in terms of coming up with a good
wavefunction for the ground state. The FQHE has $\un
>1 $ flux quanta per electron. This will be seen to frustrate
the search for a ground state.

\subsection{What  makes FQHE so
hard?}  The usual approach of considering interactions
perturbatively fails when the number of flux  quanta per particle
$>1$.
 Consider
one free electron.

\begin{eqnarray} H_0 &=& {({\bf p} + e{\bf A})^2 \over 2m}\ \ \
\ \ m\mbox{ is the bare mass}\\ e{\bf
A}&=&{eB\over2}(y,-x)=-{1\over 2l^2}\hat{z}\times {\bf r}\\ l^2&=&
{ \hbar \over eB}\ \ \ \ \ l\ \mbox { magnetic length}\\ \nabla
\times e{\bf A} &=& -eB
\end{eqnarray}
Note  $B$ points down the $z$-axis. If you do not like this
complain to the person who defined the electron charge to be $-e$
or  $z=x+iy$. Setting $\hbar =c=1$,

\begin{eqnarray} H_0&=& {\etab^2 \over 2ml^4}\\ \etab &=& {1 \over
2} {\bf r} +l^2 \hat{z}\times {\bf p}\\ \left[ \eta_{x}\ ,
 \eta_{y} \right] &=&  il^2  \\
  E &=& (n
+ {1\over 2}) {eB\over m}.
\end{eqnarray}
One calls $\etab$  the {\em cyclotron coordinate}, even though its
two components form a  conjugate pair. This is why the hamiltonian
describes a harmonic oscillator whose energy levels are called
Landau Levels (LL). Of special interest is the Lowest Landau Level
(LLL) which has $n=0$. In this state  $\langle\eta^2
\rangle_{n=0}= l^2$.

Now  a one-dimensional oscillator spectrum for a two dimensional
problem implies the LL's must be degenerate.
 The degeneracy is due to another
canonical pair  that does not enter the hamiltonian, called the
{\em guiding center coordinates} ${\bf R}$ defined by
\begin{eqnarray}
{\bf R}&=& {1 \over 2} {\bf r} -l^2 \hat{z}\times {\bf p}\\ \left[
R_{x}\ , R_{y} \right] &=& - il^2
\end{eqnarray}
Note that the role of $\hbar$ is played by $l^2$ and that  \beq
{\bf r}={\bf R}+{\bf \eta}.\eeq   which plays the role of phase
space and the degeneracy of the LLL is
 \beq D = {L^2 \over "h"} = {L^2
\over 2 \pi l^2} = { eBL^2 \over 2\pi \hbar} ={\Phi \over \Phi_0}
\eeq where $\Phi_0$ is the flux quantum. The {\bf inverse filling
factor } is \beq \nu^{-1} = \mbox{flux quanta  or LLL states per
electron} ={eB\over 2\pi n\hbar}\eeq

When $\nu^{-1}>1$ the noninteracting problem is macroscopically
degenerate since there are more LLL states than electrons. However
when $\nu^{-1}>1$ has the special values where the steps cross the
straight line,  we know excellent approximations to ground states
and low lying excitations. At these points   \beq \sigma_{xy }=
{ne\over B} = {e^2 \over 2\pi \hbar} {p \over 2ps+1} \eeq which
means that at these special points  \beq { eB\over 2\pi n\hbar } =
{ \mbox{flux quanta density}\over \mbox{particle density}} = {
\mbox{LLL states}\over \mbox{particles}}={2ps+1 \over p} \eeq

\beq \mbox{\bf flux quanta per particle} = \mbox{\bf states per
particle} = 2s + {1\over p} \eeq

Thus for example at  $\nu =2/5$ there are $5/2$ states per
particle in the LLL and the  $V=0$ problem is macroscopically
degenerate. While this is also true when there are 2.500001 states
per electron,
 we will see that at these special values, a
single ground state built out of LLL  states will emerge in
accordance with the expectation that  {\em as $m\to 0$ and $eB/m
\to \infty$,  a sensible low energy  theory built out of the LLL
states must exist!} Let us see how this happens and what happens
as we move along the plateaus.

\subsubsection{Getting to know the LLL}
The LLL condition $ a_{     \eta}     |LLL\rangle =0$ gives \beq
\psi = e^{-|z|^2/4l^2} f(z)     \eeq where $z=x+iy$. A basis for
$\psi$  is \beq     \psi_m(z) = z^m\ e^{-|z|^2/4l^2}\ \ \ \ \ \ \
m=0,1,\ldots \eeq The Gaussian is usually suppressed. The state
has $L_z=m$.

If  $\nu =1$, ( one electron per LLL state ) there is a unique
noninteracting ground state (which may then be perturbed by
standard means) \beq \chi_1 =\prod_{i<j} (z_i -z_j)\cdot Gaussian
= Det \left|
\begin{array}{cccc}
  z_{1}^{0} & z_{1}^{1} & z_{1}^{2} & .... \\
  z_{2}^{0} & z_{2}^{1} & z_{2}^{2} & ...\\
  .... & .... & .... &....
\end{array}
\right|\cdot Gaussian\eeq

For $\nu <1$, we want to work entirely within the LLL. If in
$H=T+V$ we set  $T$ equal to a constant (  $eB/2m$ per particle),
all the action is in $V$. Why is this a problem if $V$ is a
function of just coordinates? In terms of the density   \beq \rho
(\br ) =\sum_j \delta ( \br -\br_j)\eeq
  \begin{eqnarray} V&=&{1 \over 2} \int
d^2r \int d^2 r' \rho (\br ) v( \br -\br' ) \rho  (\br' )\\ &=& {1
\over 2}\sum_{\bq }\rho (\bq  )v(q) \rho (-\bq )\\ \rho (q) &=&
\sum_j e^{-i\qr_j}\end{eqnarray}

       The point is that if one sets $T= eB/2m$, the LLL value,
one must project the operator ${\bf r}$ to the LLL. The
coordinates $x$ and $y$ which commute in the full Hilbert space,
no longer commute in the LLL.

\subsubsection{     Projection to the LLL}

    Let ${\cal P}$ denote
projection to the LLL. Then \beq {\cal P}: \ {\bf r}={\bf R} +
\etab \Rightarrow  {\bf R} .\eeq The projected components do not
commute: \beq \left[ R_{x}\ , R_{y} \right] = - i{l^2}\ \ \
\mbox{or}\ \ \ \ \ \ \  \left[ {z}\ , \bar{z} \right] =  -2{l^2} \
\ \ \mbox{in the LLL}\eeq As for the densities \begin{eqnarray}
{\cal P}&:&\ e^{-i\qr}\Rightarrow \langle e^{-i{\bf q} \cdot
\etab} \rangle_{LLL}e^{-i\qR}=
e^{-q^2l^2/4}e^{-i\qR}\end{eqnarray} Thus the projected problem is
defined by
\begin{eqnarray}
\bar{\bar{H}}&=& {1\over2}\sum_{\bq}e^{-q^2l^2/2}\bar{\bar{
\rho}}(\bq )v(q)\bar{\bar {\rho}}(-\bq )\label{energy} \\ \barro
&=& \sum_j e^{-i{\qR}_j}
\end{eqnarray}
I shall refer to $\bar{\bar {\rho}}$ as the  projected density,
although it is actually the Magnetic Translation Operator and
differs from the  density by a factor $e^{-q^2l^2/4}$. (This
explains the gaussian in Eqn.(  \ref{energy})). The $\bar{\bar
{\rho}}$  obey the Girvin-MacDonald-Platzman (GMP) \cite{GMP}
algebra: \beq \left[ \bar{\bar{\rho}}({\bf q}) ,
\bar{\bar{\rho}}({\bf q'}) \right] =2i \sin \left[ { ({\bf q\times
q'})\ l^2 \over 2}\right] \bar{\bar{\rho} }({\bf q+q'}) \eeq

There is no small parameter in $\bar{\bar{H}}$ and the overall
energy scale is set by $v(q)$.
 How is one to find the
  ground state?

\subsection{Laughlin's answer}

 For $\nu =1/(2s+1)$
Laughlin proposed a trial  state: \beq \Psi_{ {1\over (2s+1)}}=
\prod_{i=2}^{N}\prod_{j<i}(z_i-z_j)^{2s+1}\exp (- \sum_i
|z_i|^2/4l^2)\eeq

Let us now contemplate its many virtues.
\begin{itemize}
  \item It lies in the LLL. (Analytic function times gaussian)
  \item It has definite angular momentum (homogeneous in z's)
  \item It obeys the Pauli principle ($2s+1 $ is odd, spin is assumed fully polarized)
  \item It has  the right filling.\\
  Proof 1: Let us refer to the wavefunction as a function of $z_N$
  with all the others fixed as
   $\Psi(z_N)$. Consider a finite sample and
  drag  $z_N$  around it.
  \beq \mbox{\#  of zeros of}\  \Psi(z_N)=
   {\mbox{Aharonov Bohm phase change}\over 2\pi}={ \Phi\over \Phi_0} \eeq
   But we know \beq
   { \Phi\over \Phi_0}=
   N\nu^{-1}\eeq

  Thus   $\Psi_{1/2s+1}(z_N) $ must be   a polynomial of degree
  $N(2s+1)$ in $z_N$ and it is.\\
  Proof 2: Again we consider a finite droplet.
  Consider any single particle wavefunction in the LLL with angular momentum $m$.
  \begin{eqnarray}
  |\psi_m|^2 &=& |z|^{2m}e^{-|z|^2/2l^2}\\
  \pi z_{max}^{2}&=& 2\pi m l^2\\
  {\Phi \ \mbox{enclosed in }\pi z_{max}^{2}\over \Phi_0} &=& {2\pi m l^2B\over (2\pi /e)}=m
  \end{eqnarray}
Thus the single particle state of  $L_z=m$ has a size (which is
sharply defined at large $m$)  such that it encloses $m$ flux
quanta. Now the largest value of $m$ for any coordinate in
  $$ \Psi_{    {1\over (2s+1)}}=
\prod_{i=2}^{N}\prod_{j<i}(z_i-z_j)^{2s+1}\exp (- \sum_i
|z_i|^2/4l^2) $$ is  $N(2s+1)$. Thus the droplet
   encloses $2s+1$ flux quanta per
  particle.
\item
Has great correlations and no wasted  zeros. Halperin observed
  \cite{halperin1} that the Laughlin function has no wasted zeros in
the following  sense.  Given that $\psi (z_N)$  has to have $N
(2s+1)$ zeros per particle (given analyticity and the filling
fraction) only $N$ of these had to lie on other electrons by the
Pauli principle. But they all lie on other electrons, thereby
keeping them away from each other very effectively, producing a
low potential energy.
\item It is not enough to know what $\Psi$ is, we need to know what
 kind of state it represents. Laughlin gave the following answer:
  it is an {\em incompressible fluid}, which means a fluid that
  abhors density changes. Whereas a Fermi gas would increase (decrease)  its
  density globally when compressed (decompressed), an
  incompressible fluid is wedded to a certain density
  and would first show no response and then
  suddenly nucleate a localized region of different density. (Just
  the way a Type II superconductor, in which magnetic field is not
  welcome, will allow it to enter in quantized units in a region
  that turns normal.) Laughlin deduced this result based on the
  following
  plasma analogy.
Note that all averages in the  ground state are found using the
measure \beq \langle \Omega \rangle = \int \prod_id^2z_i\Psi^*
\Omega \Psi/\int \prod_id^2z_i\Psi^*  \Psi\eeq
 \begin{eqnarray} \Psi^* \Psi & =&e^{-"E/kT"}\\
 {1\over " kT"}&=&2(2s+1)\\
 "E"&=& \!\! \!\! \sum_{i< j}-\ln
 |z_i-z_j| + \!\sum_i \! {\pi n \over 2} |z_i|^2
 \end{eqnarray}
 where I have used $1/l^2 =eB=2\pi n/\nu$.
This describes a plasma  with $"1/kT"=2(2s+1)$, particles of unit
charge interacting with a log potential and in a neutralizing
background given by second term.  The  plasma is known to hate
density changes away from that of the background because of the
log potential.

 Laughlin also provided the wavefunction for a state with a
 quasihole, a state with a charge deficit. If one imagines
 inserting a tiny flux tube at a point $z_0$ and slowly increasing
 the flux to one quantum, one must, by gauge invariance, return to
 an eigenstate of $H$, and each particle must undergo a $2\pi$
 phase shift as it goes around $z_0$. This and  analyticity imply the ansatz
    \beq \Psi_{qh}=     \prod_i     (z_i-z_0)
\Psi_{2s+1}\eeq      This is a      {\em quasihole}. (There is a
more complicated state with a quasiparticle.)  The prefactor is a
{\em vortex} at $z_0$, which is an analytic zero at $z_0$  for
every coordinate. It denotes a hole near $z_0$ with charge deficit
$1/(2s+1)$ (in electronic units) as can be seen using Laughlin's
wavefunction and the plasma analogy in which there  is now an
extra term in energy ${1\over 2s+1}\sum_i\ln |z_i-z_0|$. This
positive charge embedded in the plasma is immediately screened by
an equal negative quasihole charge.

  A second proof of quasihole charge depends only on the state being
  gapped and
incompressible. As the  flux quantum $\Phi_0$, is inserted, the
charge  driven out  to infinity is given by integrating the radial
current density $j$  produced by the Hall response to the induced
azimuthal $E$ field:
\begin{eqnarray} Q&=& -\int j(r,t) 2\pi r dt = -\sig  \int E\ 2\pi
r \ dt \nonumber \\ &=& -\sig \int {d\Phi \over dt} dt = -\Phi_0
\sig\\ &=& -{2\pi \hbar \over e}\ {e^2 \over 2\pi \hbar} \ \nu =
-e\nu
\end{eqnarray}
Thus fractional charge is due to fractional $\sig$ and not {\it
vice versa.}
\end{itemize}

Finally the plateaux are produced when you go off the magic
fractions  because the change in density appears in the form of
quasiparticles (holes)  which get localized, as per  standard
localization mythology in $d=2$.

While a fairly comprehensive understanding of the fundamentals
FQHE was provided by Laughlin, the subsequent years brought  many
phenomena that needed explanation. For example Girvin, MacDonald
and Platzman \cite{GMP} worked out the dispersion relation for a
magnetoexciton in which has the quantum numbers of a quasiparticle
and hole which are however not infinitely separated. Their work
showed a roton minimum just as in superfluids.

We can ask how there is a Hall current at $\nu=1/2$ if $e^*=0$,
i.e., the quasiparticles are neutral.

But our primary question next is this: what are the wavefunctions
for fractions of the form $\n$? The corresponding Laughlin
wavefunction with $2s+1 \to 2s+1/p$ does not obey the Pauli
principle. Jain gave us trial states for these fractions and also
explained why they are natural, in terms of objects called {\em
composite fermions} (CF).

 \section{Composite Fermions}
It was pointed out by Leinaas and Myrheim   \cite{leinaas} that in
$d=2$ one could have particles (dubbed anyons by Wilczek
\cite{leinaas}) that suffered a phase change $e^{i\theta}$ upon
exchange, with $\theta =0 \ ,\pi$ corresponding to bosons and
fermions respectively. To do this one takes a fermion and drives
through its center a point  flux tube. If this contains an
even/odd number of flux quanta, the composite particle one gets is
a fermion/boson. (One may ask how adding a flux quantum is going
to make any difference since it will not show up as one goes fully
around it. Consider two particles, $A$ at  the origin and $B$ at
$x=R$. To exchange them we rotate B by $\pi$ around A and shift
both to the right by $R$. )

Jain exploited this idea as follows. Consider $\nu^{-1}=2s+1/p$,
where each particle sees $2s+1/p$ flux quanta of external $B$ per
particle. Let us now convert our electrons into CF by attaching
$2s$ point flux quanta pointing opposite to $B$.  Now we invoke
the idea used by Laughlin {\em et al}  \cite{fetter} and argue
that {\em on average} the CF's see $1/p$ flux quanta per particle
and fill up exactly $p$ LL's. This gives the following trail state
at mean field level:

  \beq
\Psi_{p/2ps+1} = \prod_{i<j} \left|{(z_i -z_j)\over
|z_i-z_j|}\right|^{2s}\cdot \chi_p (z,\bar{z})\eeq  where $\chi_p$
is the CF wavefunction with $p$-filled LL's  and the prefactor
takes you back to electrons. Jain however improved this ansatz in
two ways and proposed: \beq \psi_{p/2ps+1} = {\cal P} \prod_{i<j}
 (z_i -z_j)^{2s}\cdot
    \chi_p (z,\bar{z}))\eeq

\begin{itemize}
  \item He replaces flux tubes by vortex attachment:
  $\prod_{i<j}    (z_i -z_j)^{2s}$.
  \item  He does a projection
to LLL using ${\cal P}$: \beq {\cal P} : \bar{z} \to 2l^2
{\partial \over
\partial z}\ \ \ \ \
\mbox{(Recall $\left[ z,\bar{z}\right] =-2l^2$.)}  \eeq
\item At $p=1$,\ $\chi (z, \bar{z}) =\prod (z_i-z_j)$, and  we do not need ${\cal P}$
 to  get back Laughlin's answer. At $p>1$ we have concrete
 expression for $\Psi$ in terms of electron coordinates.
\end{itemize}
The Jastrow factor \beq J(2s) = \prod_{i<j} (z_i -z_j)^{2s} \eeq
describes $2s$-fold vortices on particles. This has two effects.

First, when one CF goes around a loop, it effectively sees
$\nu^{-1*} = 2s+{1\over p}-2s ={1\over p}$ flux quanta per
enclosed particle since the phase change due to encircling
vortices attached to CF's neutralizes $2s$ of the flux quanta per
particle of the   Aharanov - Bohm phase of the external field.

{\em Thus while degeneracy of the noninteracting problem is
present for any $\nu <1$, at the Jain fractions one can beat it by
thinking in terms of composite fermions.} As we move off the Jain
fractions, the incremental  CF (particles or holes) get localized,
giving the plateaus.

Next, the vortices reduce $e$ down to \beq  e^*=1-{2ps \over
2ps+1}={1 \over 2ps+1}\eeq

The idea that electrons bind to vortices, first pointed by
Halperin\cite{halperin1}, has been greatly emphasized by  Read
\cite{read2} who also give a physical picture to explain this
binding: unlike the flux tube, the vortex is a physical LLL
excitation, one which is charged and hence attracted to the
electron.

It is however  important to note that {\em  $2s$-fold vortices are
associated with electrons only in the Laughlin states. In the Jain
states this is only before projection by}  $ {\cal P} $. The
reason is that upon projection,  zeros can  move off the electrons
(as will be explained shortly)  and also get destroyed: for
example at $\nu =2/5$, after projection to the LLL (i.e., to
analytic functions) there can only be  $5/2$ zeros per electron;
one must lie on other electron as per the Pauli's principle and
only 3/2 are left to form vortices. Clearly they cannot all be on
electrons or be associated with them uniquely.

This brings us to the following question. The quasiparticle
charge, $e^*=1-{2ps\over 2ps+1}$ is robust under ${\cal P}$ since
$e^*$ is tied to $\sigma_{xy}$ which is presumed robust under
projection. However the CF   can no longer be viewed as  an
electron-vortex complex. {\em What, if any,  is the particle that
binds to the electron to bring $e$ down to $e^*$?} The hamiltonian
theory will provide an answer.

\subsubsection{ $\nu =1/2$}

The wavefunction, also called the Rezayi-Read wavefunction
\cite{rezayi-read} is \beq
 \Psi = {\cal P} \prod_{i<j}   (z_i-z_j)^2 Det \left| e^{i{\bf k}_i\cdot {\bf r}_j}\right| .\eeq
At $\nu=1/2$, $e^*=0$ since the double vortex has charge $=-1$.
Read  \cite{read2} gave the following argument for the dipole
moment of this neutral CF. Consider
 \beq e^{i{\bf k\cdot r}} = \exp {i \over 2}
({k\bar{z}+ \bar{k}z})\ \ \  k= k_x+ik_y\eeq
   Since
$e^{il^2k{\partial \over
\partial z}} $ causes   $z \to z+ikl^2   $ in the Jastrow factor producing the  dipole moment
 $d^*=kl^2$. The energy to separate the  vortex from electron
(coulomb attraction) becomes kinetic energy of the CF. The leading
term is quadratic in separation i.e., $k$,  and defines an $m^*$.

While this picture has many merits and will be regained in an
operator version later on, there are some subtleties often
neglected.

\begin{itemize}
  \item {\em Every} $z_i$ gets translated so that $(z_i-z_j)\to
  (z_i-z_j+ik_i-ik_j)$.
  \item All this is  for  one assignment of $k$'s. Upon
  antisymmetrization, we  can't say where the zeros will be
  and there need not be a simple relation between the
  location of the electrons and the zeros in $\Psi_{LLL}$.
  \item Upon projection, vortices not only get moved around, they
  are reduced in number: at $\nu=1/2$ there are 2 vortices per
  electron (before projection there were three zeros
  per particle: two in the Jastrow factor and one in the determinant).
  One is on electrons due to the Pauli principle and the other is a {\em
  single}
  vortex.
  \item The value of dipole moment $d^*$ is sensitive to wavefunction
  unlike $\sig$: it is zero before projection (since vortices sit
  on the electrons) and reaches some value $d^*$ in the LLL, presumably $d^*=kl^2$.
   Where is one to look for this moment and what does it really mean?
  The hamiltonian
  approach will show how the dipole is described in an operator
  treatment.
\end{itemize}

The enormous success of the trial  wavefunctions notwithstanding,
the following challenges exist and are addressed by the
hamiltonian approach.

\begin{itemize}
  \item Separate high and low energies at the  cyclotron scale  and LLL.
  \item For the latter,  obtain a limit independent of $m$, and an
  energy
  scale set by $v(q)$, especially for $1/m^*$.
  \item  Find the partner to the electron that turns it
  into a CF.
  \item Obtain  the right  $e^* =1/(2ps+1)$, $d^*$, and
  $\mu^*=e/2m$, the magnetic moment predicted by  Simon,
  Stern and Halperin  \cite{SSH}.
 \item Explain who carries Hall current if CF don't. (At $\nu=1/2$, $e^*=0$.)
  \item Analyze unequal time correlations,
  $T>0$ correlations and  disorder.
\end{itemize}

\section{Hamiltonian Theory I- Chern-Simons Approach}   The aim here is to start
with the electronic hamiltonian and try to reach,  by a sequence
of transformations, exact or approximate,  fair or foul, a
description of the final quasiparticles, the CF's. So we begin
with  \beq H=\sum_i {({\bf p}_i + e{\bf A} ({\bf r}_i ))^2\over
2m}+V \eeq The next step is to attach flux tubes to electrons by
the following Chern-Simons (CS) gauge transformation  \cite{CS},
introduced into the FQHE work by  Lopez and Fradkin  \cite{lopez}
in the functional approach. In the hamiltonian version one trades
the electronic wavefunction $\Psi_{e}$ for $\Psi_{CS}$ defined as
follows :
\begin{eqnarray}
    \Psi_e &=&     \prod_{i<j} {(z_i - z_j) ^{2s}\over |z_i
-z_j|^{2s}}    \Psi_{\rm CS}\equiv   \exp
(2is\sum_{i<j}\phi_{ij})\ \Psi_{\rm CS}.\label{eq-phase}\\ H_{CS}
&=& \sum_i {({\bf p}_{i} + e{\bf A} ({\bf r}_i )+     {\bf
a}_{cs}({\bf r}_i ))^2 \over 2m} + V
\end{eqnarray}
In $H_{CS}$ there appears a CS gauge field, ${\bf a}_{cs}$, that
comes from the action of ${\bf p}$ on the prefactor, which is the
phase of the Jastrow factor:

\begin{eqnarray}
  {\bf a}_{cs}({\bf r}_i) &=&2s \nabla \sum_{j\ne i}\phi_{ij}\\
  \oint {\bf a}_{cs}({\bf r}_i)\cdot d{\bf
r}_i &=& 2s\oint \sum_{j\ne i}\nabla \phi_{ij}\cdot d{\bf r}_i
\\ &=& 4\pi s\mbox{\# enclosed}\\
       \nabla \times {{\bf a}_{cs}} &=&
4\pi s \rho
\end{eqnarray}

The above equations show that even though ${\bf a}_{cs}$ is a
gradient, it has a curl since $\phi_{ij}$ is multivalued. The curl
is readily calculated by the use of Stokes' theorem. The idea of
flux attachment was to cancel part of the applied field on
average. To this end we separate ${\bf a}_{cs}$ and $\rho$ into
average and fluctuating
  parts:\beq
    \nabla \times \langle {{\bf a}_{cs}}\rangle +    \nabla
\times :{{\bf a}_{cs}}:=    4\pi s n+     4\pi s :\rho :\eeq This
gives
\begin{eqnarray}
 H_{CS} &=& \sum_i {(    {\bf p} +e{\bf A}+\langle a_{cs}
\rangle+    :{\bf a}_{cs}:)_{i}^{2}      \over 2m} +  V \nonumber
\\ &=&\sum_i {(    \Pib +    :{\bf a}_{cs}:)^{2}_{i}\over
2m}+V
\\      \Pib &=&      {\bf p} + e{\bf A}+\langle {\bf a}_{cs} \rangle
\end{eqnarray}
\begin{eqnarray}
 \nabla \times
(e{\bf A}+\langle {\bf a}_{cs} \rangle ) & =&      -eB+4\pi ns\\
 &=&-{eB\over 2ps+1}\equiv       -e
B^*      \ \ \ \ \ \  ( A^*= {A \over 2ps+1})
\end{eqnarray}

     Here is the good news from above
       the picture derived by Fradkin and Lopez.
\begin{itemize}
  \item If we ignore $:{\bf a}_{cs}:$ and $V$, the composite
fermions see $1/p$ flux quanta each (since $2s+1/p\to 1/p$) and
have  a unique ground state $\chi_p$ of $p$ filled LL's. One can
go on to  include the neglected terms perturbatively. Excitations
are given by pushing fermions into higher CF LL's.
\item  At  mean-field, the CF wavefunction $\chi_p$, transformed
   back to
  electrons is \beq
  \Psi_e =     \prod_{i<j}\left(   {z_i-z_j \over |z_i-z_j|}\right)^{2s}    \chi_p
  (z,\bar{z})\eeq
  (See  Rajaraman and Sondhi  \cite{LLLcurrents} for a way to  get the entire
    Jastrow factor at the mean-field level, by
  introducing a {\em complex} gauge field.)
\item Fluctuations at one loop  give the square of the wavefunction
  and get rid of the
  $|z_i-z_j|$ but only for Laughlin like states.
  \item The cyclotron mode appears with the right residue.
\end{itemize}
     But there is also some bad news.
\begin{itemize}
  \item If we excite a fermion from level $p$ to $p+1$, the energy
  cost (activation gap) of the particle-hole pair is
   $\Delta = eB^*/m$ plus corrections due to neglected terms.
   The dependence on $m$ is not good. We want $\Delta \simeq e^2/\varepsilon l$.\\
   Jain does
  not have this problem: he does not use   $H_{CS}$ or $\chi_p$
  or its excitations directly.
 For him the CS picture is a step  towards  getting      electronic
  wavefunctions for the ground and excited states by attaching the Jastrow
  factor and projecting.
  The energy gap is computed  as the difference in  $\langle \bar{\bar{V}}\rangle $
  between  the  ground and excited  electronic wavefunctions.
  \item Between cyclotron mode ($eB/m$) and the LLL excitations (${\cal O} (e^2/\varepsilon l))$ there are many spurious modes
  once again due to the presence of $m$.

  \item So far we have attached flux tubes and not vortices. Consequently the
  CF have electric charge unity and not $1/(2ps+1)$.
\end{itemize}

\subsubsection{The case of   $\nu=1/2$}

Some remarkable predictions were made and verified at $\nu=1/2$.
Kalmeyer and Zhang  \cite{kz} discussed it briefly and a very
exhaustive study was made by Halperin, Lee and Read (HLR)
\cite{HLR}.

 At $\nu
=1/2$, $p=\infty$, $e^*=0$ and $\Pib \to {\bf p}$ and \beq H_{CS}
= \sum_i {(    \bp +      :{\bf a}_{cs}:)^{2}_{i}\over 2m}+V\eeq

HLR used the  Random Phase Approximation (RPA) and made the
following predictions that {\em
 did not depend on $m$}.
 \begin{itemize}
  \item A fermi surface should exist with  $k_F$ determined by $n$.
  \item At  $B^*=B-B_{1/2}$  CF should exhibit a
  bending  with   radius       $R={k_F\over eB^*}$.
  This was verified by  Kang {\em et al}  \cite{kang},
   Goldman {\em et al}  \cite{goldman} and Smet {\em et al}
    \cite{smet}.
  \item When an acoustic  wave  is coupled the electronic system,
  it must undergo a velocity shift  and an attentuation
  described by
  \beq
  {\delta v_s \over v_s}-{i\kappa \over q}={\alpha^2/2 \over
  1+i\sigma_{xx}(q)/\sigma_m}
  \eeq
where
  $ \alpha$  is a piezoelectric  constant,
   $ v_s$ is the sound velocity,
  $\kappa $ describes the  attenuation,
$\sigma_{xx}={e^2\over 8\pi \hbar}\ {q\over k_F}$ and
   $\sigma_m ={ v_s\varepsilon \over 2\pi}$.
   Theory fits the  experiments of Willett {\em at al} \cite{willett} with a $\sigma_m$ that is about
   five times larger than expected.
   \item They predicted divergences in $m^*$. These do not affect
   bosonic correlations (e.g., density-density) as shown by Kim {\em et al}  \cite{kim}.
   \end{itemize}

\subsubsection{Bosonic CS Theory}

 There is an appealing  CS theory of Zhang,Hansson and
Kivelson  \cite{zhk}  for Laughlin fractions, obtained by
converting electrons  to {\em composite bosons} in zero (mean)
field upon attaching $2s+1$ flux tubes. This has many attractions:
e.g., a Landau Ginsburg theory (see also Girvin, MacDonald
\cite{gm}, Read \cite{read}) with $FQHE \leftrightarrow
\mbox{superfluidity}$. However it too has problems due to the
unwanted dependence on $m$. Finally, unlike CF, which could be
free  in a first  approximation, composite bosons had to be
interacting for a stable ground state. Lack of time and space
prevent  further discussion of this alluring alternative.

\section{     Hamiltonian theory-II}
   I now turn to the extension of the CS theory that Murthy and I
   made in a series of papers  \cite{prl}, in order to cure it of some of its
   problems.
  We began by modifying the CS hamiltonian as follows.
   \begin{eqnarray}
H & =& \sum_i {({\Pib}_{i} +  :{\bf a}_{cs}: +      {\bf a} )^2
\over 2m} + V  \\ 0&=&     a      |physical \rangle
\end{eqnarray}
where ${     \bf a}$      is a transverse vector field. Since
 $d=2$,  it has only one component at each ${\bf q}$ which we
denote by $a(\bq )$. We introduce a conjugate variable $P(\bq )$:
\beq \left[ a({\bf q}), P({\bf -q}')\right] = i \delta_{\bf qq'}
\eeq and define a longitudinal vector field ${\bf P (q)}$  in
terms of it.  What is going on? First note that we are now
operating in a bigger Hilbert space. On vectors that are
annihilated by $a$, $H$ is the same as $H_{CS}$. Since $a$
commutes with $H$, such eigenstates of $H$ and $a$ will exist and
we focus on only those with zero eigenvalue for $a$. This is a
trick adapted  from the work of Bohm and Pines \cite{bohm-pines}.

Why do we do this? We do this to fight fire with fire: ${\bf
a}_{cs} $ is a transverse gauge field, but it is a complicated
function of the particle coordinates.  On the other hand ${\bf a}$
is an independent transverse field. {\em Let us shift ${\bf a}$ by
$-{\bf a}_{cs}$ so as to eliminate  ${\bf a}_{cs}$} from $H$. To
shift $a$ we  use the exponential of $P$: \beq U = \exp \left[
\sum_q P (q) {4\pi i s\over q} \rho ( -q)\right] .\eeq The
resulting hamiltonian and constraint on physical states are
\begin{eqnarray} H & =& \sum_i {({\Pib}_{i}  +      {\bf a}(\br_i) +
4\pi s P(\br_i))^2      \over 2m} +  V  \\ 0 &=& (q\      a (q) -
4 \pi s \rho (q)) |physical \rangle
\end{eqnarray}

The presence of ${\bf P}$ in $H$ is due to the fact that $U$
affects the particle sector as well. The nice thing is that $H$ is
written in terms of {\em independent} Fermi and Bose fields. Let
us therefore focus on the quadratic part of $H$:

\begin{eqnarray}
H &=& \sum_i {{\bf \Pi}^{2}_{i}\over 2m} + \underbrace{{n\over
2m}\sum_q (     a^2(q) + (4\pi s      P     (q))^2 )}_{\Huge
\sum_q      A^{\dag}A \omega_0 {2ps\over 2ps+1}     }\nonumber \\
&+&\mbox{V }+ :\rho: \mbox{term} + {\bf j \cdot      A      } \
\mbox{term}\label{hfree} \\ & &\nonumber
\\ 0 &=& (q      a       - 4\pi s \rho ) |physical \rangle
\end{eqnarray}

For those who want to know, here is how one gets the second term.
 \begin{eqnarray*} \lefteqn{\sum_i ( {\bf a}(\br_i) + 4\pi
s {\bf P}(\br_i))^2 =}\\
 & &\int d^2r \rho (\br )( {\bf a}(\br ) +
4\pi s {\bf P}(\br ))^2= n\int d^2r ( {\bf a}(\br ) + 4\pi s P(\br
))^2 +{\cal O}:\rho :\nonumber  \\ &=& n\sum_{\bq} (\bf{a}^2 (\bq
)+(4\pi s {\bf P }(\bq ))^2+{\cal O}:\rho :
\end{eqnarray*}

How does this help? Focus on first two terms in Eqn.
(\ref{hfree}). The ground state is a product: the  fermions are in
$\chi_p$, the state with $p$ -filled CF Landau levels, the
oscillators, whose frequency is close to   $\omega_0$, are in
their  ground state and

\begin{eqnarray}
\Psi_{CS}&=& e^{- C  \sum_q      a^2      } \chi_p \ \ \ \ \ \ \ \
\mbox{$C$ is some constant}
\\ &=& \exp [-C' \sum_q :\rho (q): {1 \over q^2} :\rho (-q):]
\chi_p \nonumber =     |Jastrow|     \chi_p\\ \Psi_{e}&=&
\mbox{Jastrow } \cdot     \chi_p
\end{eqnarray}
where  $C'$is another  constant and we have set $a(q) \simeq \rho
(q)/q$ so that
\begin{eqnarray*}
\lefteqn{C'  \sum_q :\rho (q) :{1 \over q^2} :\rho (-q) :\simeq}
\\ &  &\!\!\!\! \!\!\!\!\!\!\!\!\!\! \int \!\! d^2rd^2r' \sum_i s
(\delta (r-\! r_i)\! -\! n) \ln |r-r'| (\sum_j \delta (r'\! -\!
r_j)\!  -\! n)\\ &=& \sum_{i<j}2s\ln |z_i-z_j|-\sum_i
{|z_i|^2\over 4l^{2}_{v}}+\mbox{constant}\\ {1 \over l_{v}^{2}}&=&
{1 \over l^2}{2ps\over 2ps+1}
\end{eqnarray*}
Going from the CS to the electron  basis, we put in the phase of
the Jastrow factor, and  using $1/l^{2}_{v}+1/l^{*2}\ (
\mbox{from} \chi_p )=1/l^2$, we even get the right gaussian.

  Thus  we get the Laughlin and unprojected Jain
  wavefunctions.   However, if
 we had not  known   how good these wavefunctions were, we
 would not have  bet much money on them since they came
 from an uncontrolled approximation. However, given what we know, it appears
we are  on the right track.

 But we still have complaints.

\begin{itemize}
  \item The oscillators are  at $\omega_0 \ {2ps\over 2ps+1}$ and violate
  Kohn's  Theorem.
  \item The  kinetic energy scale for fermions is  still set by $1/m$.
  The potential energy V  has played no role,
   instead of dominating the LLL physics.
  \item High and low energy degrees of freedom are still mixed up.
  \item No sign of $e^*$,$d^*$, or the  moment $\mu^*=e/2m$
  predicted by Simon {\em et al} \cite{SSH}.
\end{itemize}
The solution is to  decouple oscillators and fermions. They are
coupled by the "$ j\cdot       A$"      term. Since we have an $
A^{\dag}A$ term, we want to shift the oscillators to complete the
squares using (schematically) \beq U = \exp \left[ C \sum_q j (q)
A^{\dag} (q) + h.c.\right]\eeq The transformation due to $U$ is
implemented with two approximations:
\begin{itemize}
  \item Infrared limit: \ \ $ql<<1$
  \item  RPA:\  $\sum_j e^{-i({\bf q-q'})\cdot \br_j}\simeq n\delta ({\bf
  q-q'})$
\end{itemize}

Upon decoupling, the theory takes the following form (for $s=1$)
\cite{prl}:
\begin{eqnarray}
H&=&      \sum_{\bq} \omega_0 A^{\dag}(\bq )A(\bq )+    \sum_i
{e\delta B ({\bf r}_i)\over      2m}\nonumber \\
 & & \nonumber  \\
 &+&       \sum_j {|\Pi_j |^2
\over 2m} - {1 \over 2mn}\sum_i\sum_j\Pi_{i}^{-}e^{-i\bq \cdot
({\bf r}_i-{\bf r}_j)}\Pi_{j}^{+}\nonumber  \\ &+& V(\rho )\\ & &
\nonumber  \\ J^{+}(\bq ) &=&     {e(q_x+iq_y)\over
q\sqrt{2\pi}}\omega_0 c      A(\bq )\ \ \ J^{\pm}=J_x\pm iJ_y\\
          c^2 &=& {2p\over 2p+1}\\ \rho (\bq ) &=&
\sum_je^{-i\qr_j}\left( 1-il^2 {\bq \times \Pib_i \over
1+c}+..\right )\\ & +&
     {cq\over \sqrt{8\pi}}(A (\bq ) + A^{\dag}(\bq ))\\
 & & \nonumber  \\     \chi
(\bq )&=&  \sum_je^{-i\qr_j}\left( 1+il^2 {\bq \times \Pib_j \over
c(1+c)}+..\right) \\ 0&=&\chi (\bq ) |\mbox{physical}\rangle
\end{eqnarray}
Let us digest the various terms. First consider $H_{os}$, the
oscillator part,  to which we will add a coupling between the
external potential $\Phi_{ext}$ and  the oscillator part of the
charge density: \beq \rho (\bq ) = {cq\over \sqrt{8\pi}}     (A
(\bq ) + A^{\dag}(\bq ))\eeq
      \beq H_{os}=
\sum_{\bq} \omega_0 A^{\dag}(\bq )A(\bq )+ e\Phi_{ext}
(\bq){cq\over \sqrt{8\pi}}(A (\bq ) + A^{\dag}(\bq ))\eeq Note
that the current depends only on $A$:
  \beq J^+(\bq )
= {eq_+\over q\sqrt{2\pi}}\omega_0 c A(\bq ) \eeq

\begin{itemize}
  \item Kohn's theorem is obeyed upon decoupling.
  \item The constraint $\chi$ makes no reference to
$     A$, which is thus truly decoupled. We have separated the
high and low energy physics in the infrared limit.
  \item Both $J$ and  $\sig$ are due to $   A$  alone.
    We find

 \beq \langle       A      \rangle = {ecq\over
\omega_0\sqrt{8\pi}}\Phi_{ext}\eeq
 Plugging  it
into $J (     A      )$ and dividing by $E$  we get \beq \sig =
{e^2\over 4\pi} c^2 ={e^2\over 2\pi}{p\over 2p+1}\eeq {\em Thus
the oscillators carry the  Hall current, known to be non LLL
effect}: \beq \sig \simeq \underbrace{{1\over m}}_{\mbox{from
$J$}} \cdot \underbrace{{1 \over (eB/m)}}_{\mbox{ energy denom
}}\simeq m^0 \eeq

\item The matrix elements of the oscillator part of $\rho$ saturates
the sum rule \beq
 \int_{0}^{\infty} S(q,\omega )\omega {d\omega }={q^2n\over
2m}\eeq \beq
 S(q,\omega )= \sum_N |\langle N|\rho (q) |0\rangle
|^2 \delta (\omega -E_N)\eeq where  \beq
\rho (q) =
\bar{\bar{\rho}}(q) + "A+A^{\dag}"\mbox{part}\eeq
\end{itemize}

     {\em  The LLL (CF) part  (due to $\bar{\bar{\rho}}(q)$)  must therefore go as
$q^4$}. Now we turn to the particle sector.

\begin{itemize}
  \item The particles  have the magnetic moment ${e\over 2m}$
   deduced by  Simon {\em et al} \cite{SSH}.
  \item The kinetic term becomes
  \beq \sum_j {|\Pi_j |^2 \over
2m} - {1 \over 2mn}\sum_i\sum_j\Pi_{i}^{-}e^{-i\bq \cdot ({\bf
r}_i-{\bf r}_j)}\Pi_{j}^{+}\label{28} \eeq      The  $i=j$ term
when combined with the first,  renormalizes the mass
    \beq {1\over m^*}={1\over m}(1-{\sum_q\over n})     \eeq
As oscillators are added and decoupled, $1/m^*$ decreases. We can
make it go away if we choose $\sum_q = \mbox{number of
oscillators} = n$, the number of particles. Let us then make this
choice.
\end{itemize}
The $i \ne j$ terms in Eqn. {\ref{28})  can be shown to be
convertible to a short range piece \cite{prl}. The infrared low
energy hamiltonian and constraint become
        \beq
 H = \bar{V} = {1 \over 2} \sum_q \rho (q) v(q) \rho (-q)\eeq

\begin{eqnarray}
{\rho}  &=&     \underbrace{ \sum_j e^{-i{\bf q \cdot r }_j}\left[
1 - {il_{}^{2}\over (1+c) } {\bf q}\times {\bf \Pi}_{j}
\right]}_{\bar{\rho}} +  ("      A+A^{\dag}") \nonumber
\\     \bar{ \chi} &=&  \sum_j e^{-i{\bf q \cdot r}_j}\left[ 1 +
{il_{}^{2}\over c(1+c) }  {\bf q}\times {\bf \Pi}_{j}+\ldots
\right]     \nonumber \\ c^2 &=& {2ps\over 2ps+1}\nonumber
\end{eqnarray}

Hereafter we will   set oscillators in their ground state and
$A=A^{\dag}=0$.

Murthy and I used this low $q$ theory to study gaps  \cite{gaps}
for potentials that vanished quickly at large $q$. There were
however some conceptual problems that bothered me. For example the
 constraint algebra does not close:     $\left[ \chi_{\alpha} ,
\chi_{\beta} \right] \ne f_{\alpha \beta \gamma}\chi_{\gamma}$.
The error was of course due to higher order terms in $q$. It did
not make sense that two constraints each annihilated the physical
states but their commutator did not. (The charge algebra did not
close either, but this at least did not pose the same problem.)
Secondly, it was not clear what the constraints stood for. I
proposed the following remedy \cite{allq}. {\em Let us assume that
the power series for charge and constraint mark the beginning of
exponential series.} Thus we assume that they extend to:

\begin{eqnarray}
\bar{\bar{\rho}} &=& \! \sum_j \exp (-i{\bf  q} \! \cdot \! ({\bf
r}_j\! - \! {l^2\over 1+c}\hat{\bf z}\times {\bf \Pi}_j ))\equiv
\sum_je^{-i{\bf q\cdot R_{ej}}}\nonumber     \\ \bar{\bar{\chi}}
&=& \! \sum_j \exp (-i{\bf  q}\! \cdot  \!({\bf  r}_j \!+
\!{l^2\over c+c^2}\hat{\bf z}\times {\bf \Pi}_j ))\equiv \! \sum_j
e^{-i{\bf q\cdot R_{vj}}}\nonumber
\end{eqnarray}
Note that  {\em   ${\bf R}_e\      \mbox{and}\      {\bf R}_{v} $
were fully determined by the small $ql$ theory}.
  Suddenly a lot of things fall into place.
\begin{itemize}
\item Consider  ${\bR}_e$ and its  commutation relations
\begin{eqnarray}
{\bf R}_e &=& {\bf r} -{l^2\over (1+c)}\hat{\bf z}\times {\bf \Pi
}\label{re}, \\ \left[ R_{ex}\ , R_{ey} \right] &=& - il^2,\
\end{eqnarray}
which correspond to  the electron guiding center, {\em but
represented  in the CF basis}. In particular, it is written in
terms of the velocity operator $\Pib$, which sees a weaker field
$A^*$ and will lead to a nondegenerate ground state with $p$
filled CF Landau levels in Hartree-Fock calculations.
\item The commutation relations of
\begin{eqnarray}
{\bf R}_v &=& {\bf r} +{l^2\over c(1+c)}\hat{\bf z}\times {\bf \Pi
}\label{rv}\\ \left[ R_{vx}\ , R_{vy} \right] &=&  il^2/c^2, \ \ \
\end{eqnarray}
correspond to the guiding center coordinates of a particle of
charge $-c^2 = -2ps/(2ps+1)$. This is the right charge for an
object that should bind with the electron to form the CF. We shall
refer to it as the vortex even though, as we have seen, in the
Jain case there aren't enough zeros to associate a $2s$-fold
vortex with each electron. \item The two sets of coordinates
commute:   \beq \left[ {\bf R}_e\ , {\bf R}_{v} \right] = 0.
\label{evcomm} \eeq

\item Will the vortex pair with the electron? Figure
  \ref{cfcoordinates} shows that the separation between the two is
 of order  $\Pi$.  Since there will be  terms in $H$ that
associate an energy with $\Pi$, the two will bind. Finally since
$H$ is just $V$ in the new basis, the binding has its origin in
the electrostatic interaction between electrons, as envisaged by
Read.
\item The constraint $\bar{\bar{\chi}} =0$ states that the vortices will have
no density fluctuations. In other words, one member of every CF
(the vortex) obeys a collective constraint on density.

\begin{figure}[t]
\begin{center}
\includegraphics[width=.5\textwidth]{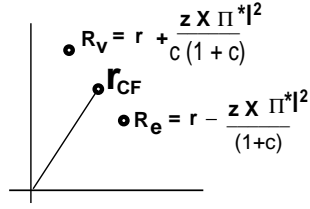}
\end{center}
\vspace*{-1in} \caption[]{ Structure of the Composite Fermion}
\label{cfcoordinates}
\end{figure}

\item The density and constraint obey the following  algebras:
\begin{eqnarray}
 \left[ \bar{\bar{\rho}}(q) , \bar{\bar{\rho}}(q') \right] &=& 2i \sin
\left[ {l_{}^{2} ({\bf q\times q'}) \over 2}\right]
\bar{\bar{\rho} }(q+q')\nonumber \\ & & \nonumber \\    \left[
\bar{\bar{\chi}}(q) , \bar{\bar{\chi}}(q') \right] &=& -2i \sin
\left[ {l_{}^{2} ({\bf q\times q'}) \over 2c^2}\right]
\bar{\bar{\chi}} (q+q') \nonumber \\ & &
\\ \left[     \bar{\bar{\chi}}(q) ,   \bar{\bar{\rho}}(q') \right] &=&      0
\nonumber
 \end{eqnarray}
\end{itemize}
Pasquier and Haldane  \cite{pasquier}, working in the LLL, studied
bosons at $\nu=1$ and obtained an algebra that coincides with the
above for the case $c=1$.

  Here is where we stand:
\begin{eqnarray}
H &=& {1 \over 2} \sum     \bar{\bar{\rho}}(q)\ v(q)e^{-(ql)^2/2}\
\bar{\bar{\rho}} (q)\\
   0&=&  \left[ H ,    \bar{\bar{\chi}}      \right]  \\
    \bar{\bar{\chi}}      &=&0 \label{natural}\end{eqnarray}
   As in Yang-Mills, the constraints form an
  algebra that commutes with $H$.

Could we perhaps ignore the constraint $\bar{\bar{\chi}} =0$?
 Consider $\nu =1/2$ where
 \beq \bar{\bar{\rho}}({\bf q})  =  \sum_j  e^{-i{\bf q \cdot
r}_j} \left[       1      -
  {il_{}^{2}\over      2     }
{\bf q} \times {\bf p}_{j} +\cdots \right]
 \eeq
The fermion has unit charge and half the expected dipole moment.
Presumably
    $\bar{\bar{\chi}}$      should  send  $e \to e^*$ and $d\to d^*$.
    Read  \cite{read3}
showed in a     conserving approximation     that
$\bar{\bar{\chi}}$ generates a gauge field whose longitudinal part
makes $e^*=0$ and the surviving dipoles interact via the
transverse field whose propagator is peaked  at $\omega \simeq
q^3$ and thus expected to matter only at $T=0$ and at the Fermi
surface.

Murthy \cite{maggm} computed magnetoexciton dispersion in a
conserving approximation. The results obeyed Kohn's theorem
($S(q)\simeq q^4$).

       A shortcut to $    \chi $ in the safe region     ($T>0$, and/or\
\  $\omega \ \mbox{not} \  <<< q^3$) was provided by Murthy and
myself\cite{prl,gaps}. We   used the {\em      preferred
combination }
 \beq \bar{\bar{\rho}}^p =     \bar{\bar{\rho
}} -      c^2    \bar{\bar{\chi}} \eeq
  equivalent to  $    \bar{\bar{\rho }}$       and  {\em weakly gauge invariant}:

\beq \left[    \bar{\bar{\chi}} \ , \bar{\bar{\rho}}^p\right]
\simeq    \bar{\bar{\chi}}. \eeq      Clearly
$H(\bar{\bar{\rho}}^p) $ is also weakly gauge invartaint.
 Consider the series
\beq \bar{\bar{\rho}}^p = \sum_j e^{-i{\bf q \cdot r}_j}\!\!
\left( \!      {1 \over 2ps+1}     \!  -     {i l_{}^{2}      }
q\times {\bf \Pi}_{j}  \! +{0} \cdot \left( q\times {\bf \Pi}_{j}
\right)^2 \! + \cdots \right) \label{rostar} \eeq Amazingly this
 single combination  yields   the correct $e^*$ and $d^*$
and can be verified to have  $q^2$ matrix elements between
Hartree-Fock states, in compliance with Kohn's theorem. It also
makes physical sense: $\bar{\bar{\rho}}^p =\bar{\bar{\rho }} - c^2
\bar{\bar{\chi}}$
  adds the  charge of electrons to
 that of correlation hole, namely
      $-c^2=-{2ps\over 2ps+1}$    and describes the correlated entity, the CF.

      Henceforth we will work with $H(\bar{\bar{\rho}}^p)$. Its
      significance is the following. In the constrained space
      $\bar{\bar{\chi}} =0$, there are many equivalent hamiltonians. In the HF
      approximation, these are not equivalent and
      $H(\bar{\bar{\rho}}^p)$ best approximates (within HF states and in the infrared)
      the true hamiltonian between true eigenstates. In contrast
      to a variational calculation where one searches among states
      for an optimal one, here the HF states are the same for a
      class of hamiltonians (where  $\bar{\bar{\chi}}$ is introduced into $H$
      in any fashion as long as rotational invariance holds) and
      we seek the best hamiltonian: $\bar{\bar{\rho}}^p$ encodes the fact
       that every electron
 is accompanied by a correlation hole of some sort which leads to
 a certain $e^*$, $d^*$ and obeys Kohn's theorem ($q^2$ matrix element in the
 LLL).

However when gauge invariance (constraints) are crucial, one must
not use $H(\bar{\bar{\rho}}^p)$ but revert to the conserving
approximation. Here is an example.

    \subsubsection{     Compressibility at $\nu =1/2$}
 A question that came up when the dipolar form of the charge
 operator was derived was this.
 If $\bar{\bar{\rho }}^p$ is dipolar,  is the system compressible?
Early approximate calculations suggested that compressibility
vanishes as $q^2$ due to the extra $q$ in the dipole moment.

 Halperin and Stern  \cite{HS1} showed through  an example that dipolar fermions can
 be compressible if $H$ has the right symmetries. These are
 part of the  symmetries generated by $\bar{\bar{\chi}}$.  In particular as $q
 \to 0$, $\bar{\bar{\chi}} (\bq )\simeq \sum_je^{-i\qr_j}$. Its action is
 \beq \delta \br_j=\left[ \bar{\bar{\chi}} , \br_j\right] =0 \eeq
 \beq \delta
 \bp_r = \left[ \bar{\bar{\chi}} ({\bf q}_0) , \bp_j\right]
 \simeq \bq_0\ \ \ \mbox{as} \ \bq_0 \to 0\eeq
 This symmetry of $H$ is called       K-invariance     and had been noticed earlier by Haldane.
 It makes the Fermi surface very squishy and produces a response
 whose singularity cancels the $q^2$. The detailed work of   Stern {\em et al}  \cite{HS1}, showed that finite compressibility is ensured if the CF
 has a Landau parameter $F_1=-1$.

 Compressibility at $\nu=1/2$ was also confirmed by Read  \cite{read3}  within  the
 conserving approximation
 and by
 D.H.Lee  \cite{dh}.

That  the correct compressibility obtains only if the
 constraint $\bar{\bar{\chi}} =0$ is correctly imposed makes physical
 sense: if one end of every dipolar fermion (the vortex) is actually part of an
 inert system with no density fluctuations (for this is what $\bar{\bar{\chi}} =0$ means)
  the other, electronic,  end (with charge
 unity) must responds with  a nonvanishing compressibility.

\subsection{      My Final Answer       }
For the rest of these lectures I will use
\begin{eqnarray}
\Huge H^p &=& {1 \over 2} \sum_q \bar{\bar{\rho}}^p(q)
\check{v}(q) \bar{\bar{\rho}}^p(-q)\\
 \bar{\bar{\rho}}(q)&=& \sum_j e^{-i{\bf q \cdot r_j}}\
\left[ {1\over 2ps+1}-il^2{\bf q \times \Pi_j}+\mbox{known series}
\right]\nonumber\\
 \check{v}(q)
&=& v(q) \ e^{-q^2l^2/2}= {2\pi e^2 e^{-ql\lambda}\over
q}e^{-q^2l^2/2}     \ \ \ \ ZDS
\end{eqnarray}
Note that $v(q)$ is the Zhang-Das Sarma potential  \cite{zds}. I
use it just to illustrate the formalism: it has a free parameter
$\lambda$ which allows one to suppress large $q$. Roughly
speaking, $\lambda$ is a measure of sample thickness in units of
$l$.

  I illustrate the
  rather unusual form
  of  $H$ we have been led to by considering the simplest case of $\nu =1/2$.
  When we square $\bar{\bar{\rho}}^p$ we get a double sum over particles whose
   diagonal part is the one particle (free field) term:
     \beq
H^{0}_{\nu =1/2}=2\sum_j \int {d^2q\over 4\pi^2} \sin^2
\left[{{\bf q \times     k     }_j l^2\over 2}\right] \check{v}(q)
\eeq
  This is not a hamiltonian of the form $k^2/2m^*$. However if the
  potential is peaked at very small $q$, we can expand the $\sin$
  and  read off an approximate $1/m^*$
   \beq  {1\over m^*}= \int {qdq d \theta \over 4\pi^2}
\left[ (\sin^2 \theta )\ (ql)^2 \right] \check{v} (q)\eeq  which
has its origin in
  electron-electron interactions. However we can do more: we have
  full $H_0$ as well as the interactions. The point I want to
  emphasize is that $H$ is not of the traditional form and that
  there is no reason it had to be.

\section{Computation of Gaps} The formalism will be illustrated
with some examples, starting with activation gaps for a fully
polarized sample. The expression for the gap is \beq \Delta_a =
\langle {\bf p} + PH | H|{\bf p}+ PH \rangle -\langle {\bf p} | H|
{\bf p} \rangle \eeq where $| {\bf p} \rangle $ stands for a HF
ground state with
 $p$-filled LL's and
 $|{\bf p}+ PH \rangle$  for the
  state with a widely separated particle-hole pair.  The
  hamiltonian is $H=H^p$. It turns out all the matrix elements can be analytically
  evaluated. Figure   \ref{PMJ}   compares the numbers so obtained with
  those obtained by Park, Meskini, and Jain  \cite{pmj} using Jain
wavefunctions. The gap formula looks the same for them,
  except that the states are not the simple ones (like $\chi_p$)  mentioned   above,
  but these
  multiplied by  the Jastrow factor and projected down to the
  LLL. On the other hand $H=V(\rho )$ is very simple in the electronic basis with
  $\rho (q) =\sum_j e^{-i{\bf q \cdot r_j}}.$

\begin{figure}[b]
\begin{center}
\includegraphics[width=.5\textwidth]{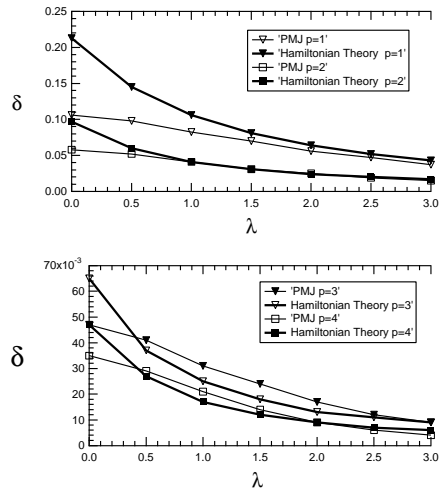}
\end{center}
\caption[]{Dimensionless activation gaps $\delta =
\Delta_a/(e^2/\varepsilon l)$ compared the work of PMJ. }
\label{PMJ}
\end{figure}

Note that for $p=3,4$, the HF answer is not necessarily above the
PMJ results, perhaps because the  PMJ results may  not be ideal
benchmarks. Indeed when I compare my results to the exact
diagonalization work of Morf {\em et al}  \cite{Morf}, I find that
the HF numbers are systematically above, as shown in Fig.
(\ref{morff}). (The $b$ parameter is like $\lambda$ and defines a
very similar potential.)

\begin{figure}[b]
\label{morff}
\begin{center}
\includegraphics[width=.5\textwidth]{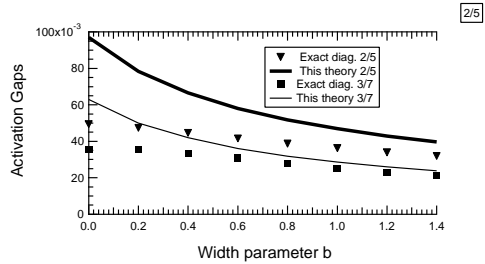}
\end{center}
\caption[]{Dimensionless gaps compared to exact diagonalization by
Morf {\em et al}. }\label{morff}
\end{figure}

I also compared the numbers to PMJ for a gaussian potential and
found almost perfect agreement except for $1/3$. In general it
appears the theory works best for potentials that are soft at
large $q$, say for $\lambda >1$.

Are CF weakly interacting? Given  that two different mass scales
control activation and polarization processes, one expects the
answer to be negative, though only the hamiltonian formalism, with
a  concrete $H$, allows us to ask this in a meaningful way. In Fig
  \ref{freeornot},  I compare the effect of turning off the interaction
on gaps.  Note that interactions seem less important for $\nu
=1/4$ and
 systematically get less important as $p$ increases.
  I do not  understand this.

\begin{figure}[b]
\begin{center}
\includegraphics[width=.5\textwidth]{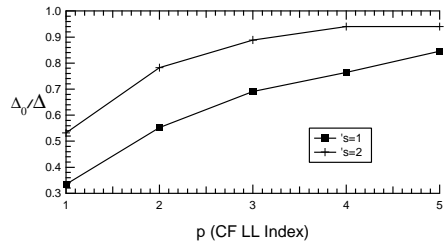}
\end{center}
\caption[]{Effect of turning off the interaction on gaps. }
\label{freeornot}
\end{figure}

I next compare the theory to the experiments of Du {\em et al}
\cite{du} in Table   \ref{du}. This is very tricky since we do not
know the exact form of interactions and have not included
disorder. The idea is to see what sort of $\lambda$ fit the data.

\begin{table}
\caption{Comparison to Du {\em et al}}
\begin{center}
\renewcommand{\arraystretch}{1.4}
\setlength\tabcolsep{5pt}
\begin{tabular}{lllll}
\hline\noalign{\smallskip} $ \nu \ \ \  $  &$B(T)$&
$\Delta_{a}^{exp} ({}^oK)$  & $\Delta_{a}^{theo} ({}^oK)$ &
$\lambda$\\ \noalign{\smallskip} \hline \noalign{\smallskip}
 1/3  &13.9 & 8.2 &$5.3 \sqrt{B(T)}/\lambda$ & 2.4  \\
 2/5 & 11.6&3 & $2.08 \sqrt{B(T)}/\lambda$ &2.4  \\
3/7  &10.8 &2 &$1.23 \sqrt{B(T)}/\lambda$ &2.0 \\ \hline
\end{tabular}
\end{center}
\label{du}
\end{table}

The value of $\lambda \simeq 2$ is roughly double what one expects
for a pure system. Additionally, the gaps, when extrapolated to
$\nu =1/2$ have a negative intercept of a few Kelvin. These
suggest that disorder is very strong and that it is not possible
to describe a disordered system by an effective $\lambda$.

Consider now the results of Pan {\em et al}  \cite{Pan} who found
that   \beq m^{nor}_{a}= {m_a\over m_e \sqrt{B(T)}}\simeq .25-.35.
\eeq
  near
$\nu =1/2$ and $1/4$. How does this rough equality of normalized
masses of fermions with two and four vortices fit in the present
theory? I find
\begin{eqnarray}
  m^{nor}_{a}
  &=& .163 \lambda \ \ (s=1)\label{nor1}\\
  &=& .175 \lambda^{5/4}\ \  (s=2)\label{nor2}
  \end{eqnarray}
  where the powers of $\lambda$  are approximate.
  The theory does explain the near equality of these masses, but
  clearly does not attribute  any fundamental significance to it since
  the answer is sensitive to the potential ($\lambda$).

\subsection{Magnetic phenomena}
So far we have assumed the spin to be fully polarized along the
applied field. Thus in the fraction $p/(2ps+1)$, the CF fill
$p$-LL with spins up. This however costs a lot of kinetic energy,
which would favor filling spin-up and down LL's equally.   If one
could vary the Zeeman coupling (by placing the sample in a tilted
field whose normal part remains fixed) one could drive the system
through various transitions. If $E(p-r,r)$ is the energy of the
state with $p-r$ up and $r$ down LL's (not including Zeeman
energy) then the transition $r\to r+1$ will take place at a field
$B^c$ given by  \beq E(p-r,r)-E(p-r-1,r+1) = g{ e  B^c\over
2m_e}{n\over p} \label{criticalB}\eeq When the energies are
computed, a strange fact, previously seen by Park and Jain,
emerges: even though the CF are not free, the energy differences
behave as if CF were free and occupied LL with a (polarization)
gap $\Delta_p$. That is,  we find \beq E(p\! -\! r,r)-E(p\! -\!
r\! -\!1,r\! +\! 1)= {n(p-2r-1 )\over p} \Delta_p
\label{gapdefine} \eeq (This would be the relation in a free
theory  since $(n/p)$ spin-up fermions of energy $(p-r-1+{1\over
2})\Delta_p$ drop to the spin-down level with energy  $ (r+{1\over
2}) \Delta_p$).

In the gapless case,  the polarization is determined  by \beq
{\cal E}_+(k_{+F}) -{\cal E}_-(k_{-F}) = g{e\over 2m_e} {B_{\perp}
\over \cos \theta} \eeq where $k_{\pm F}$ are the Fermi momenta of
up and down fermions and ${\cal E}_{\pm F}$ the corresponding
Fermi energies. Calculations show once again that the Fermi energy
differences may be fit by \beq {\cal E}_+(k_{+F}) -{\cal
E}_{-}(k_{-F})= {k_{+F}^{2}-k_{-F}^{2}\over 2m_p} \eeq where $m_p$
is  the polarization mass.

But we know CF cannot be  not free because the activation gap
$\Delta_a=eB^*/m_a$ and polarization gap  $\Delta_p=eB^*/m_p$
 are very different.   {\em Here is my argument   \cite{magnetic} that this free field behaviour for
 magnetic phenomena is accidental
 and  reflects  $d=2$ and rotational invariance.}
 Let us assume that the energy as a function of total spin $S$ has
 the form
  \beq E(S) =
E(0)+{\alpha \over 2}S^2\eeq where $\alpha$ is the inverse linear
static susceptibility. Consider the gapless case for simplicity.
When  $dn$  particles go from  spin-down to spin-up,
 \begin{eqnarray}
 dE &=&{\alpha   } \ S\  dS = {\alpha   }\  S \ ( 2 \ dn)\\
 &=& \alpha {k_{+F}^{2}- k_{-F}^{2}\over 4\pi} (2 \   dn)
 \end{eqnarray}
using  the areas of the circular Fermi seas.
  We see that $dE$  has precisely the form of the kinetic
 energy difference of particles of mass  $m_p$ given by
\beq
 {1 \over m_p} = {\alpha \over \pi}.
  \eeq
 {\em Thus $m_p$ is essentially  the static susceptibility,
  which happens to have dimensions of mass in
 $d=2$. }
With this understanding I give the calculated values

\begin{eqnarray}
{1 \over m^{(2)}_{p}} &=& { e^2 l \over \varepsilon }C_{p}^{(2)}
(\lambda )\ \ \ \
 \ \ \ \  C_{p}^{(2)} (\lambda )  = {.087 \over \lambda^{7/4}} \label{mp22}\\
{1 \over m^{(4)}_{p}} &=& { e^2 l \over \varepsilon }C_{p}^{(4)}
(\lambda ) \ \ \ \ \ \ \ \
   C_{p}^{(4)} (\lambda )={.120 \over \lambda^{7/4}}\label{mp44}
\end{eqnarray}
where the superscripts on $C$ refer to the number of vortices
attached and the exponent $7/4$ is approximate.

\subsubsection{Comparison to $T=0$ experiments}
Since magnetic transitions are controlled by ground state
energies,  perhaps disorder can be incorporated via an effective
potential. {\em I have shown  \cite{disc} that under certain
restrictive  conditions this is possible and will try to fit
theory to experiment by using one data point to extract an
effective $\lambda$ and use it  to make predictions for other
measurements made on that sample ( using simple scaling laws for
$\lambda$ if needed).}

 Kukushkin {\em et al}  \cite{kuk} vary both $n$ and a perpendicular
 $B $ to
drive the system through various $T=0$ (by extrapolation)
transitions.  I will compare the hamiltonian  theory to these
experiments by computing the  $B^c$'s at which the systems at
$1/4,2/5,3/7,4/9, $ and $1/2$ lose full polarization ($r=0$ for
gapped cases, saturation for the gapless cases) and, for $4/9$,
also  the $r=1$ transition, $|\bf{3,1}\rangle \to
|\bf{2,2}\rangle$. I fit $\lambda$ to the $\nu =3/7$ transition
$|\bf{3,0}\rangle \to |\bf{2,1}\rangle$ at $B^c=4.5 T$. The
results are in Table \ref{kuk}.

\begin{table}
\caption{Comparison to Kukushkin {\em et al}. Critical fields
based on a fit at $3/7$.The rows are ordered by the last column
which measures density.}
\begin{center}
\renewcommand{\arraystretch}{1.4}
\setlength\tabcolsep{5pt}
\begin{tabular}{lllll}
\hline\noalign{\smallskip} $ \nu \ \ \  $ & comment &  \  $B^{c}$
(exp) &\  $B^{c}$ (theo) &\ \ $\nu B^c$ (exp)\\
\noalign{\smallskip} \hline \noalign{\smallskip} 4/9 & $(3,1)\to
(2,2)$ & \ \ 2.7 T & \ \ 1.6 T & \ \  1.2\\
 2/5 & $(2,0)\to (1,1)$ &\ \ 3 T & \ \   2.65 T & \ \  1.2\\
1/4 & saturation &\ \   5.2 T &\ \  4.4 T &\ \  1.3\\ 3/7 &
$(3,0)\to (2,1)$ &\ \ 4.5 T&\ \   4.5 T & \ \    1.93\\ 4/9 &
$(4,0)\to (3,1)$ & \ \ 5.9 T & \ \  5.9 T & \ \   2.62\\ 1/2 &
saturation&\ \  9.3 T&\ \ 11.8 T&\ \  4.65\\ \hline
\end{tabular}
\end{center}
\label{kuk}
\end{table}

\section{Physics at $T>0$}
The hamiltonian formalism is particularly suited to study physics
at $T>0$ in the infinite volume limit. Given a concrete
hamiltonian, one can work out the HF energies at $T>0$
\begin{eqnarray*}
\lefteqn{ {\cal E}_{\pm}^{Z}(k)=}\\ &&\mp {1 \over 2}g \left[ { e
B\over 2m}\right] + 2\int{d^2q \over 4\pi^2}\check{v}(q)\sin^2
\left[{{\bf k \times q}l^2\over 2}\right] \\ && -4\int{d^2k' \over
4\pi^2}n^{F}_{\pm}(|k'|)\check{v}(|{\bf k -k'}|)\sin^2 \left[{{\bf
k' \times k}l^2\over 2}\right]
\end{eqnarray*}
where the superscript on ${\cal E}_{\pm}^{Z}$ reminds us it is the
total energy including the Zeeman part, the Fermi functions \beq
n^{F}_{\pm}(|k|)={1 \over \exp \left[ ({\cal E}_{\pm}^{Z}(k)-\mu
)/kT\right] + 1 }\eeq depend on the energies ${\cal
E}_{\pm}^{Z}(k)$ and the chemical potential $\mu$. At each $T$,
one must choose a $\mu$, solve for ${\cal E}_{\pm}^{Z}(k)$ till a
self-consistent answer with the right total particle density $n$
is obtained. From this one may obtain the polarization by taking
the difference of up and down densities.

The computation of $1/T_1$ is more involved.  The theory predicts
\cite{disc}
\begin{eqnarray}
1 \over T_{1} &=& -16\pi^3 k_BT\left( {K^{max}_{\nu}\over
n}\right)^2\nonumber \\ &&\! \! \! \! \! \! \! \times
 \int_{E_0}^{\infty} dE \left(
{dn^F(E)\over dE}\right)
\rho_{+}(E)\rho_{-}(E)F(k_+,k_-)\label{oneovert}\\
F&=&e^{-(k_{+}^{2}+k_{-}^{2})l^2/2}I_0(k_+k_-l^2)\\
 \rho_{\pm}(E) &=&\int{kdk\over
2\pi}\delta (E-{\cal E}_{\pm}^{Z}(k))\label{oneovert1}
\end{eqnarray}
where $E_0$  is the lowest possible energy for up spin
fermions,$k_{\pm}$ are defined by  ${\cal E}_{\pm}^{Z}(k_{\pm})=E$
and $K^{max}_{\nu}$ is the measured maximum Knight shift.
\subsubsection{Comparison to experiment}

\begin{figure}[t]
\begin{center}
\includegraphics[width=.5\textwidth]{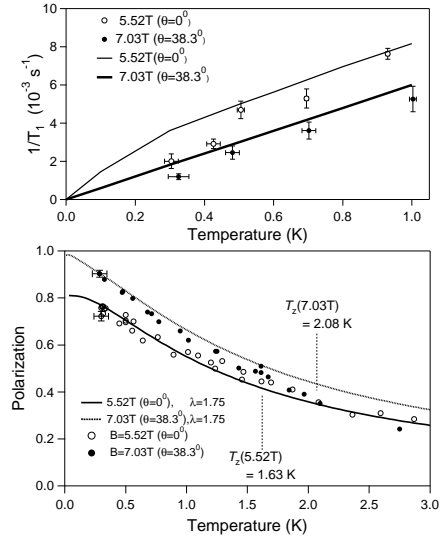}
\end{center}
\caption[]{Comparison to the work of Dementyev {\em et al}.
  The value of $\lambda$ is fit to $P$ at $300 \ mK,\  B_{\perp}=5.52\ T$
   and the  rest follows from the theory. }
\label{dem}
\end{figure}

\begin{figure}[b]
\begin{center}
\includegraphics[width=1.2\textwidth]{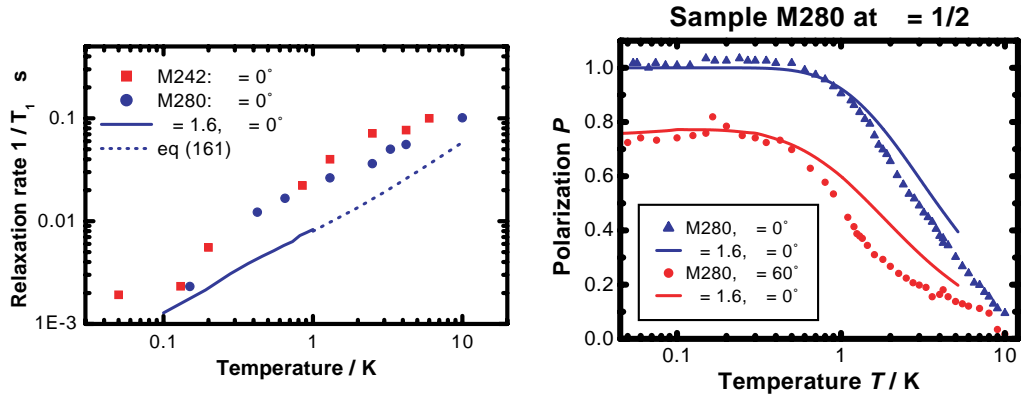}
\end{center}
\caption[]{Comparison to the work of Freytag {\em et al}.
  The value of $\lambda$ was fit to $1.6$
   and the  rest follows from the theory. }
\label{mel}
\end{figure}

We now compare to some experiments at $\nu=1/2$ and $T>0$.
 Consider first Dementyev
{\em et al}   \cite{dem}.  From their data point  $P=.75$ for
$B=B_{\perp}= 5.52 T$ at $300 \ mK$ I deduce $ \lambda =1.75. $ I
have once again  chosen  to match my HF results with the above
data point,  and see to what extent a {\em sole} parameter
$\lambda$, can
 describe  $P$ and $1/T_1$ for the
given sample at a given $B_{\perp}$, but various temperatures and
tilts. The results are shown in Fig.   \ref{dem}. Dementyev {\em
et al} had pointed out that a two parameter fit
  (mass $m$ and interaction  $J$), led to  four disjoint sets
for these four curves. Given that $H$ is neither free nor of the
standard form ($p^2/2m +V (x) $) this is to be expected. By
contrast, a single $\lambda$  is able to describe the data   since
$H$ has the right functional form.

Next I compare the results to that of Melinte {\em et al}
\cite{mel} for polarization and Freytag {\em et al}  \cite{mel}
for $1/T_1$. A value of $\lambda =1.6$ was extracted from one
polarization point  for sample M280.  Note that  a factor of two
difference in $1/T_1$. I do not have an explanation for this. If
the theory is correct, the figure implies that there are other
means of relaxation besides the one introduced into the
calculation.

\section{Summary}
Electrons in a magnetic field form degenerate Landau Levels and
for the cases we were interested in ($\nu <1$), all the electrons
could be fit into the Lowest Landau Level (LLL) with room to
spare. The macroscopic degeneracy of the noninteracting ground
state ruled out perturbative treatments. Interactions were
expected to  lift the degeneracy and produce a unique ground state
and tower of excited states with a scale set by the interactions.
How was one to  find them?

 One way out was to write down inspired trial wavefunctions in the LLL that
 had all the right properties, a trail blazed by Laughlin for $\nu =1/(2s+1)$. From
 his work we learnt also that the system was an incompressible
 fluid that allowed only localized density deviations. For example
 a vortex at $z_0$ described a quasihole with a charge deficit of
 $1/(2s+1)$.

 Jain extended the
 wavefunctions to fractions $\nu =p/(2ps+1)$. By trading an electron for
  CF which carried $2s$ units of
 flux that, on average, canceled $2s$ of the $2s+1/p$ flux quanta per
 particle,
 Jain obtained a particle  that saw $1/p$ flux quanta per
 particle, which was just right to fill $p$ LL's in the
 noninteracting case, a state we called $\chi_p$. The electronic
  wavefunction was obtained by undoing the flux attachment.
  In practice, the flux tubes were replaced by
 vortices and a projection was done to the LLL.

From these wavefunctions we learnt  \cite{jain-cf},\cite{read2}
that in the Laughlin case electrons were bound to $2s$-fold
vortices to form CF with a charge $e^*$. This was  true  in the
Jain case only before projection, which kills many zeros and also
moves them off the particles. It was however true in both cases
that the CF had a charge $e^*$.

We then asked how all these could be derived directly from the
hamiltonian. We saw how flux could be attached by the Linneas and
Myrrheim or Chern-Simons transformation  following Fradkin and
Lopez. Then the work of Murthy and myself was described. Here one
introduces additional (oscillator) degrees of freedom at the
cyclotron scale and a corresponding number of constraints $\chi$.
Placing the oscillators in the ground states gave us the modulus
of the Jastrow factor which combined with the phase in the CS
transformation to give the Laughlin and unprojected Jain
wavefunctions. These oscillators were then decoupled from the
fermions in the infrared-RPA approximation. The oscillators, now
at exactly the cyclotron frequency, were seen to carry all the
Hall current. The particles' kinetic energy could be quenched if
the number of oscillators was chosen to equal the number of
particles. In the low energy sector we were then left with just
the potential energy $V$ written in terms of the projected density
$\bar{\rho}$ (in the new basis) and a set of constraints
${\bar{\chi}}$.

We then turned to my extension of these expressions for all $ql$
(denoted by $\bar{\bar{\rho}}$ and $\bar{\bar{\chi}}$) which not
only made the problem mathematically more alluring, but captured
much of the known CF physics in operator form. In particular, one
could see a CF made of the electron and another object  which had
the charge of the $2s$-fold vortex (called vortex for simplicity).
The potential energy of electrons was seen to bind the two and to
give the CF its charge and  internal structure, as well as kinetic
and interaction energies exactly as desired. The Hilbert space of
the CF was seen to spawn two guiding center coordinates, one for
the electron and one for the vortex. The constraints prevented the
latter from having density fluctuations. Hence the  proper
implementation of constraints was needed to recover the HLR result
of finite compressibility at $\nu=1/2$: with one end of every CF
frozen, only the electronic end responded to any applied potential
 as a unit charge object.

There were two approaches to solving the theory defined by
$H=V(\bar{\bar{\rho}})$ and subject to the constraints
$\bar{\bar{\chi}}=0$. Since the $\left[ \bar{\bar{\chi}}, H\right]
=0 $ one could find an approximation that respected the constraint
algebra, as was done by Read for the gapless case or Murthy
\cite{maggm} for magnetoplasmons. This is ideal when gauge
invariance is important, typically at $T=0$, in the gapless cases,
at the Fermi surface. For the other cases, (emphasized here), we
use the one proposed by Murthy and myself: make  the replacement
$\bar{\bar{\rho}} \to \bar{\bar{\rho}} -c^2 \bar{\bar{\chi}}\equiv
\bar{\bar{\rho}}^p$ which incorporates many of the effects of the
constraints in a HF calculation and respects Kohn's theorem.

We focused on a few illustrative examples of this formalism.  We
saw that gaps could be computed for the polarized case in HF in
closed form and that they were within about $10\%$ of the
wavefunction based results or exact diagonalization results, for
potentials that were soft for large $q$, which meant roughly that
$\lambda >1$, $\lambda$ being the parameter in the Zhang-Das Sarma
potential.

We could see that CF were not free by turning off the interaction
term and observing sizeable changes in gaps. (Without a
hamiltonian such a question did not even have a meaning.) We could
understand the rough equality of the renormalized  masses observed
by Pan {\em et al}.

Turning to magnetic phenomena, we understood how polarization
phenomena could be mimicked by a free theory with a LL gap
$\Delta_p$ due a conspiracy involving $d=2$ and rotational
invariance. Magnetic transitions at $T=0$ observed by Kukushkin
{\em et al} were described by the theory given one data point from
which an effective $\lambda$ was deduced. (Heuristic arguments
suggesting that magnetic phenomena  in a disordered  system could
be described by an effective $\lambda$ was demonstrated elsewhere
\cite{disc}.)

We saw how it was possible to compute polarization and relaxation
at $\nu =1/2$ as a function of $T$ using just one data point to
extract $\lambda$. The power of this approach was evident in the
comparison to the data  of Dementyev et al: whereas a single
$\lambda =1.75$ gave a very good description of two polarization
and two relaxation rate  graphs, a fit to the data with a
 canonical mass plus interaction term
 required four disjoint sets of values. The theory was also in
 fair agreement with the  Melinte {\em at al} and Freytag
 {\em et al} data.

In summary, we understand  the FQHE in two complimentary fronts:
trial wavefunctions and hamiltonian approaches.  The former give
excellent numbers where applicable and the latter provide many
interpolating steps, insight,  and  facilitate otherwise
impossible computations  such as unequal-time correlations,
coupling to disorder or relaxation rates at $T>0$, all  in the
infinite volume  limit.

 After this brief introduction, inevitably limited and idiosyncratic,
  you should be  ready to
pursue variants of the simple FQHE system studied here. Some are
reviewed in Ref. \ref{review2} by Eisenstein (experiments on
double layer systems),
 Girvin and MacDonald (theory of the same), and  Kane and Fisher
(edge physics).
  Other possible areas that might interest you are areas are drag \cite{drag} and
 skyrmions \cite{skyrmions}.

%


\begin{thebibliography}{8.}
\addcontentsline{toc}{section}{References}
\bibitem{review1}  For a review see {\em The Quantum Hall Effect},
Edited by R.E.Prange and S.M Girvin, Springer-Verlag, 1990, T.
Chakraborty and P. Pieti\"{a}\"{i}nen, {\em The Fractional Quantum
Hall Effect: Properties of an incompressible quantum fluid},
Springer Series in Solid State Sciences, {\bf 85}, Springer-Verlag
New York, 1988, A.H. MacDonald ed., {\em Quantum Hall Effect: A
Perspective}, Kluwer, Boston, 1989. A. Karlhede, S. A. Kivelson
and S. L. Sondhi, ``The Quantum Hall Effect: The Article'', in
V.J. Emery ed.,``Correlated Electron Systems'', World Scientific,
Singapore (1993)\label{review1}
\bibitem{review2}  {\it Perspectives in
Quantum Hall Effects}, Edited by Sankar Das Sarma and Aron Pinczuk
( Wiley, New York, 1997).\label{review2}
\bibitem{olle}{\it Composite
Fermions}, Edited by Olle Heinonen, World Scientific, Singapore,
(1998).\label{olle}
\bibitem{baron} K. von Klitzing, G. Dorda and M. Pepper,
{\em Phys. Rev. Lett.}, {\bf 45}, 494, (1980)
\bibitem{fqhe-ex} D.Tsui, H.Stromer and A.Gossard, Phys. Rev. Lett.,
{\bf 48}, 1599, (1982). (FQHE).
\bibitem{laugh} R. B. Laughlin, Phys. Rev. Lett, {\bf 50}, 1395, (1983). Needs no explanation.
\bibitem{jain-cf} J. Jain, Phys. Rev. Lett., {\bf 63}, 199, (1989).
 For the latest summary see  J.K. Jain and R.
Kamilla, in Ref(  \ref{olle}).
\bibitem{GMP} S.M.Girvin, A.H. MacDonald and P. Platzman, Phys. Rev.
{\bf B33}, 2481, (1986).
\bibitem{halperin1} B.I. Halperin, Helv. Phys. Acta {\bf 56}, 75,
(1983).
 B.I. Halperin,  Phys. Rev. Lett.,  {\bf 52},
1583, (1984).
\bibitem{leinaas} J.M.Leinaas and J.Myrheim, {\it Nuovo Cimento} {\bf
37B}, 1 (1977).  F. Wilczek, Phys. Rev. Lett. {\bf 48}, 1144,
(1982).
\bibitem{fetter} R. L. Laughlin, Phys. Rev. Lett., {\bf 60}, 2677 , (1988). See
also A.L.Fetter, C.B. Hanna, and R.B. Laughlin, Phys. Rev. {\bf
B39}, 9679, (1989), {\em ibid} {\bf 43}, 309, (1991).
\bibitem{read2} N.Read Semi. Cond. Sci. Tech., {\bf 9}, 1859, (1994),
and Surface Science, {\bf 361/362}, 7, (1995).
\bibitem{rezayi-read} E. Rezayi and N.Read, Phys. Rev. Lett,{\bf 72},
900, (1994).
\bibitem{SSH}  A.Stern, and B.I.Halperin, Phys. Rev. {\bf
B54}, 11114 (1996). See the review on modifications of RPA by
S.Simon, J. Phys. Cond. Mat., {\bf 48}, 10127 (1996).
\bibitem{CS} S. Deser, R. Jackiw and S. Templeton, Phys. Lett., B139,
371,  {\em Field theories in condensed matter physics}, E.
Fradkin, Addison-Wesley, Reading MA 1991.
\bibitem{lopez} A. Lopez and E.Fradkin, Phys. Rev. {\bf B 44}, 5246,
(1991), {\em ibid} {\bf 47}, 7080, (1993), Phys. Rev. Lett., {\bf
69}, 2126, (1992).
\bibitem{LLLcurrents} R. Rajaraman and S. L. Sondhi, Mod. Phys. Lett. B {\bf 8}, 1065
(1994).
\bibitem{kz} V.Kalmeyer and S. C. Zhang, Phys. Rev.{\bf B46}, 9889,
(1992).
\bibitem{HLR} B. I. Halperin, P.A.Lee and N.Read, Phys. Rev. {\bf B47},
7312, (1993).
\bibitem{kang} W.Kang, H.L. St\"{o}rmer, L.N. Pfeiffer, K.W. Baldwin
and K.W. West, {\em Phys. Rev. Lett.}, {\bf 71}, 3850, (1993).
\bibitem{goldman} V.J. Goldman, B. Su and J. K. Jain,
{\em Phys. Rev. Lett.}, {\bf 72}, 2065, (1994).
\bibitem{smet} J.H.Smet, D.Weiss, R.H. Blick, G. Lutjering and K.
von Klizting, {\em Phys. Rev. Lett.}, {\bf 77}, 2272, (1996).
\bibitem{willett} R.L. Willett, M .A. Paalanen, R.R.Ruel, K.W.West,
L.N.Pfeiffer, and D.J.Bishop, Phys. Rev. Lett. {\bf 54}, 112
(1990) For a review see R. Willett, Advances in Physics., {\bf
46}, 447, (1997).
\bibitem{kim} Y.B. Kim, A. Furusaki and P.A. Lee,Phys. Rev.{\bf B50},
 17917,
(1994).
\bibitem{zhk}S.C. Zhang, H.Hansson and S.Kivelson, Phys. Rev. Lett.,
{\bf 62}, 82, (1989). See also D.-H. Lee and S.-C. Zhang, Phys.
Rev. Lett. {\bf 66}, 1220 (1991),   S.C. Zhang, Int. J. Mod.
Phys., {\bf B6}, 25, (1992) and
 C. L. Kane, S. Kivelson, D.H. Lee and S.C.Zhang, Phys.
Rev.  {\bf B 43 }, 3255 (1991).
\bibitem{gm} S.M. Girvin and A.H. MacDonald,
Phys. Rev. Lett. {\bf 58}, 1252, (1987). See also S. M. Girvin in
Ref.   \ref{review1} for work pointing to Chern-Simons theories.
\bibitem{read}N. Read, Phys. Rev. Lett. {\bf 62}, 86,
(1989).
\bibitem{prl} R.Shankar and G. Murthy, Phys. Rev. Lett., {\bf 79},
4437, ( 1997). For details see   "Field Theory of the FQHE", in
Heinonen's  book (Ref. (  \ref{olle}) or  cond-mat 9802244.
\bibitem{bohm-pines} D. Bohm and D. Pines, Phys. Rev. {\bf 92}, 609,
(1953).
\bibitem{gaps} G. Murthy and R.Shankar,Phys. Rev. {\bf B 59 }, 12260, (1999),
  G.Murthy, R.Shankar, K.Park, and J.K.Jain,
   Phys. Rev. {\bf B58}, 13263, (1998).
\bibitem{allq}  R.Shankar, Phys. Rev. Lett., {\bf
83}, 2382, (1999) cond-mat 9903064.
\bibitem{pasquier}
V. Pasquier and F.D.M.Haldane Nucl. Phys. {\bf 516}, 719,(1998).
\bibitem{read3} N. Read, Phys. Rev. {\bf B58}, 16262, (1998).
\bibitem{maggm}
 G. Murthy, cond mat. 9903187.
\bibitem{HS1} B.I.Halperin and A.Stern, Phys. Rev. Lett.,  {\bf 80}, .5457
(1998); G. Murthy and R.Shankar, {\em ibid} , 5458 (1998). A.
Stern, B.I. Halperin, F. von Oppen and S.Simon, Phys. Rev.,
{\bf59},12547, (1999).
 cond-mat
9812135.
\bibitem{dh}  D.H. Lee, Phys. Rev. Lett., {\bf 80}, 4745 (1998).
\bibitem{zds} F.C. Zhang
and S. Das Sarma, Phys. Rev. {\b B 33}, 2908, (1986).
\bibitem{pmj} K. Park, N.Meskini,  and J.K. Jain, J. Phys. Condensed Matter, {\bf 11}, 7283, (1999).
Has a  response to   R.Morf, Phys. Rev. Lett. {\bf 83}, 1485,
(1999).\label{pmj}
\bibitem{Morf} R.H. Morf, N. d'Ambrumenil and S. Das Sarma
(to be published).
\bibitem{du} R.R.Du, A.S. Yeh, H.L. St\"{o}rmer, D.C. Tsui, L. N. Pfeiffer, K.W.Baldwin
 and K.W. West, Phys. Rev. Lett., {\bf 70}, 2944, (1993).
 \bibitem{Pan} W. Pan, H.L. St\"{o}rmer, D.C.Tsui, L.N. Pfeiffer, K.W. Baldwin
and K.W. West,
 Phys. Rev., {\bf B61}, R5101, (2000).
\bibitem{magnetic}  R.Shankar, Phys. Rev. Lett., {\bf
84},3946, ( 2000) cond-mat 9911288.
 \bibitem{disc} R.Shankar, Phys. Rev {\bf B 63}, 085322, (2001),
 cond-mat 0009361.
\bibitem{kuk} I.V.Kukushkin, K. v.
Klitzing and K. Eberl, Phys. Rev. Lett. {\bf 82}, 3665, (1999.)
\bibitem{dem} A.E. Dementyev, N.N. Kuzma, P. Khandelwal, S.E. Barrett,
 L.N. Pfeiffer, and K. W. West, Phys. Rev. Lett., {\bf 83}, 5074, (1999). cond-mat/9907280.
\bibitem{mel} S.Melinte, N.Freytag, M. Horvatic, C. Berthier, L.P. Levy,
 V. Bayot and M. Shayegan, Phys. Rev. Lett., {\bf 84}, 354, (2000).
  cond-mat/9908098 are the source of $P(\theta =0)$. The rest are
  from N. Freytag, L. Levy, M. Horvatic, C. Berthier, and M. Shayegan,
  unpublished.
  \bibitem{drag} For a recent review see A.G. Rojo,
     Electron-drag effects in coupled electron systems
     J. Phys. Cond. Mat.,  { \bf 11 (5)}: R31-R52, (1999).
     For some very early work see G.L. Zheng and A.H.  Macdonald,
      Phys. Rev {\bf B 48}, 8203, (1993). For work on FQHE see
      for example I. Ussishkin I and A.  Stern
    Phys. rev. , {\bf  56 (7)}, 4013, (1997).
 \bibitem{skyrmions} For a review see S. M. Girvin, `The Quantum Hall Effect: Novel Excitations and
Broken Symmetries,' 120 pp. Les Houches Lecture Notes, in
Topological Aspects of Low Dimensional Systems, ed. by Alain
Comtet, Thierry Jolicoeur, Stephane Ouvry and Francois David,
(Springer-Verlag, Berlin and Les Editions de Physique, Les Ulis,
2000, ISBN: 3-540-66909-4). cond-mat/9907002,  The first  papers
were S.\,L. Sondhi, A. Karlhede, S.\,A. Kivelson,
        E.\,H. Rezayi,  Phys. Rev. B {\bf 47}, 16419 (1993).
  H.A. Fertig, L. Brey, R. C\^ot\'e, A.H. MacDonald,
A. Karlhede, and S.L. Sondhi, Phys. Rev. B {\bf 55}, 10671 (1997),
and  S. E.  Barrett {\em et al.},
       Phys. Rev. Lett., {\bf 74}, 5112 (1995).
\end{thebibliography}
\end{document}